# Indoor optical fiber eavesdropping approach and its avoidance

HAIQING HAO,[1] ZHONGWANG PANG,[1,2] GUAN WANG,[1,2] AND BO WANG [1,2,*]

[1]*State Key Laboratory of Precision Measurement Technology and Instruments, Department of Precision Instrument, Tsinghua University, Beijing 100084, China*
[2]*Key Laboratory of Photonic Control Technology (Tsinghua University), Ministry of Education, Beijing, 100084, China*
**bo.wang@tsinghua.edu.cn*

**Abstract:** The optical fiber network has become a worldwide infrastructure. In addition to the basic functions in telecommunication, its sensing ability has attracted more and more attention. In this paper, we discuss the risk of household fiber being used for eavesdropping and demonstrate its performance in the lab. Using a 3-meter tail fiber in front of the household optical modem, voices of normal human speech can be eavesdropped by a laser interferometer and recovered 1.1 km away. The detection distance limit and system noise are analyzed quantitatively. We also give some practical ways to prevent eavesdropping through household fiber.

## 1. Introduction

In recent years, optical fiber networks are widely deployed all over the world, which not only facilitates data transmission but also provides an opportunity to obtain additional information. These applications of optical fiber networks, including earthquake detection [1-6], urban traffic flow monitoring [7-10], underground geological structure exploration [11-15], etc., have positive impacts on people's production and life. However, it also brings some potential security problems, which should be considered carefully.

Optical fibers are sensitive to environmental pressure variations, which could be induced by acoustic waves. Devices based on this feature are widely used in sound detection, such as fiber-optic hydrophones [16, 17]. V. Grishachev points out the risk of voice information leak through optical fiber networks [18] and discusses its feasibility in the office scenario [19]. According to the current layout mode of fiber to the home (FTTH), tail fiber up to several meters will be installed in residents' homes. For these indoor fibers, sound signals could be modulated onto the light wave transmitted therein, allowing other people to eavesdrop and recover them at remote places along the fiber link. Meanwhile, the original communication function of the fiber will not be influenced by using wavelength division multiplexer (WDM). Therefore, eavesdropping may be carried out in a secret way.

In this paper, we propose an eavesdropping scheme using indoor optical fiber and demonstrate it in the laboratory. The system is based on the Mach-Zehnder heterodyne interferometer. We couple the eavesdropping system into the optical fiber link 1.1 kilometer away from the eavesdropping target, voices of normal human speech (50~80 dB) can be eavesdropped with a 3-meter indoor tail fiber. The system noise and eavesdropping ability are analyzed. Finally, we discuss the measures to avoid the risk of eavesdropping.

## 2. Experimental Setup

Fig. 1(a) shows the diagram of the eavesdropping scheme. People talking in the building and the sound wave will change the optical length of the indoor tail fibers. Eavesdroppers could detect corresponding phase changes and recover them to sound signals outside the building.

It’s not required to install any additional equipment in the resident's home. What they should do is to find the correct optical fiber and couple it with the eavesdropping system.

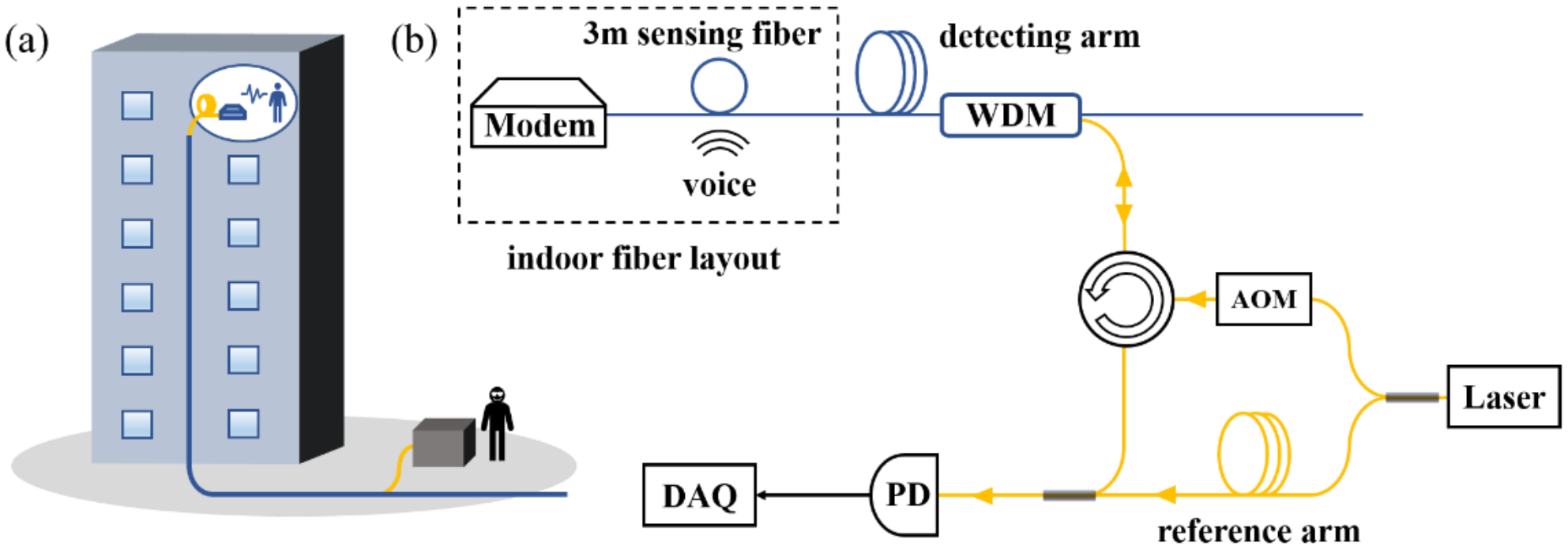

Fig. 1(a) Diagram of the eavesdropping scheme. (b) Heterodyne interferometer-based eavesdropping system, Laser (NKT Koheras BASIK Laser Module), AOM (Acousto-Optic Modulator), PD (Photodiode), DAQ (Data Acquisition), WDM (Wavelength Division Multiplexer).

Fig. 1(b) shows the indoor optical fiber eavesdropping system demonstrated in the laboratory. Blue lines stand for the communication optical cable, and yellow lines denote the inserted eavesdropping element. The dotted box represents the target resident's room, where a 3-meter tail fiber connected to an optical modem is settled. During the experiment, other parts of the system are physically isolated from the dotted part to ensure not being acted on by sound waves.

In the experiment, the cw laser is split into two beams. One beam goes through the reference arm and reaches the photodiode (PD) directly. The other beam is frequency shifted by an 80 MHz acousto-optic modulator, goes through the circulator and couples into the detecting arm via a WDM. The detecting arm includes the 3-meter tail fiber and 1.1 km fiber spool. The spool is used to simulate the scenario that the eavesdropping happens 1.1 km away from the house. At the input connector of the optical modem, the laser beam is partially reflected by the PC-type fiber end face, with reflectivity of around 4%. The reflected beam is interfered with the reference beam and detected by the PD. To make the interferometer balanced, the reference arm is twice the length of the detecting arm.

The laser source is an NKT Koheras BASIK module with linewidth of around 100 Hz. The PD detected heterodyne signal is sampled by a 400 kS/s data acquisition system (DAQ). The sound signal is loaded on the 3-meter tail fiber of the detecting arm, which is marked as sensing fiber in Fig. 1(b). The intensity of the sound is measured by a decibel meter.

At the position of PD, the reference beam can be written as:

$$E_{Reference}\left(t\right)=E_0\exp i\left[\omega_0 t+\varphi_{Laser}\left(t\right)+\varphi_{F-R}(t)+\varphi_0\right], \tag{1}$$

where $E_0$ is the amplitude and $\omega_0$ is the center frequency of the laser, $\varphi_{Laser}\left(t\right)$ is the phase variation of the laser source, $\varphi_{F-R}\left(t\right)$ is the phase variation induced by the reference arm and $\varphi_0$ is the initial phase. The reflected beam can be described as:

$$E_{Reflect}\left(t\right)=\alpha E_0\exp i\left[\left(\omega_0+\omega_{AOM}\right)\left(t-\tau_0\right)+\varphi_{Laser}\left(t-\tau_0\right)+\varphi_{F-D}\left(t\right)+\varphi_{Voice}\left(t\right)+\varphi_0\right], \tag{2}$$

where $\alpha$ is a scale factor, $\tau_0$ is the time delay between two beams, $\varphi_{F\text{-}D}\left(t\right)$ is the phase variation induced by the detecting arm and $\varphi_{Voice}\left(t\right)$ is the phase variation caused by sound waves on the 3-meter tail fiber, which is our eavesdropping target. The heterodyne signal detected by PD can be written as:

$$
\begin{aligned}
I(t) &= \left[E_{Reference}(t)+E_{Reflect}(t)\right]\cdot\left[E_{Reference}(t)+E_{Reflect}(t)\right] \\
&= \left(1+\alpha^2\right)E_0^{\,2} \\
&\quad +2\alpha E_0^{\,2}\cos\left[\omega_{AOM}t-\left(\omega_0+\omega_{AOM}\right)\tau_0+\varphi_{Voice}(t)+\varphi_{Laser}(t-\tau_0)-\varphi_{Laser}(t)+\varphi_{F-D}(t)-\varphi_{F-R}(t)\right].
\end{aligned}
\tag{3}
$$

Through DAQ and IQ demodulation, the phase information can be extracted and described as:

$$
\phi(t)=\varphi_{Voice}(t)+\varphi_{Laser}(t-\tau_0)-\varphi_{Laser}(t)+\varphi_{F-D}(t)-\varphi_{F-R}(t). \tag{4}
$$

## 3. Theoretical Analysis

From equation (4), it is clear that the phase variation caused by sound is mixed with two basic noise terms. $\varphi_{F-D}(t)-\varphi_{F-R}(t)$ in equation (4) illustrate the effect of the thermal noise induced by fiber links, and $\varphi_{Laser}(t-\tau_0)-\varphi_{Laser}(t)$ shows the noise from the frequency fluctuation of the laser source. An estimation of these noise level is necessary, because the signal-to-noise ratio (SNR) of the converted phase signal is deeply correlated with the system's background noise level [20]. We use the root-mean-square (RMS) of phase variation to evaluate the noise level, which represents the average phase change caused by noise. Considering that only phase variations within the audible frequency range (from 100 Hz to 10 kHz) are wanted, variations out of this range can be filtered out, without affecting the voice quality. Therefore, we only discuss the characteristics of noise in this frequency range.

The power spectral density (PSD) of the fiber thermal phase noise can be described as [21, 22]:

$$
S(f)=\left(\frac{2\pi}{\lambda}n\right)^2\frac{2k_BTL\gamma_0}{3\pi KA}\frac{1}{f}, \tag{5}
$$

where $K$ and $A$ are the bulk modulus and the cross-sectional area of the fiber, and their product maintains when replacing the fiber's coat. $\lambda$ is the wavelength of the laser, $k_B$ is the Boltzmann constant, $\gamma_0$ is the loss angle, $n$ is the refraction index of the fiber core, $L$ is the fiber length and $T$ is the temperature.

The RMS phase variation can be calculated as:

$$
\sqrt{\left\langle\Delta\phi^2\right\rangle}=\left(\int_{f_L}^{f_H}S(f)\,df\right)^{1/2}, \tag{6}
$$

where the upper and lower limits $f_H=10kHz$ and $f_L=100Hz$ contain the frequency range of the human voice.

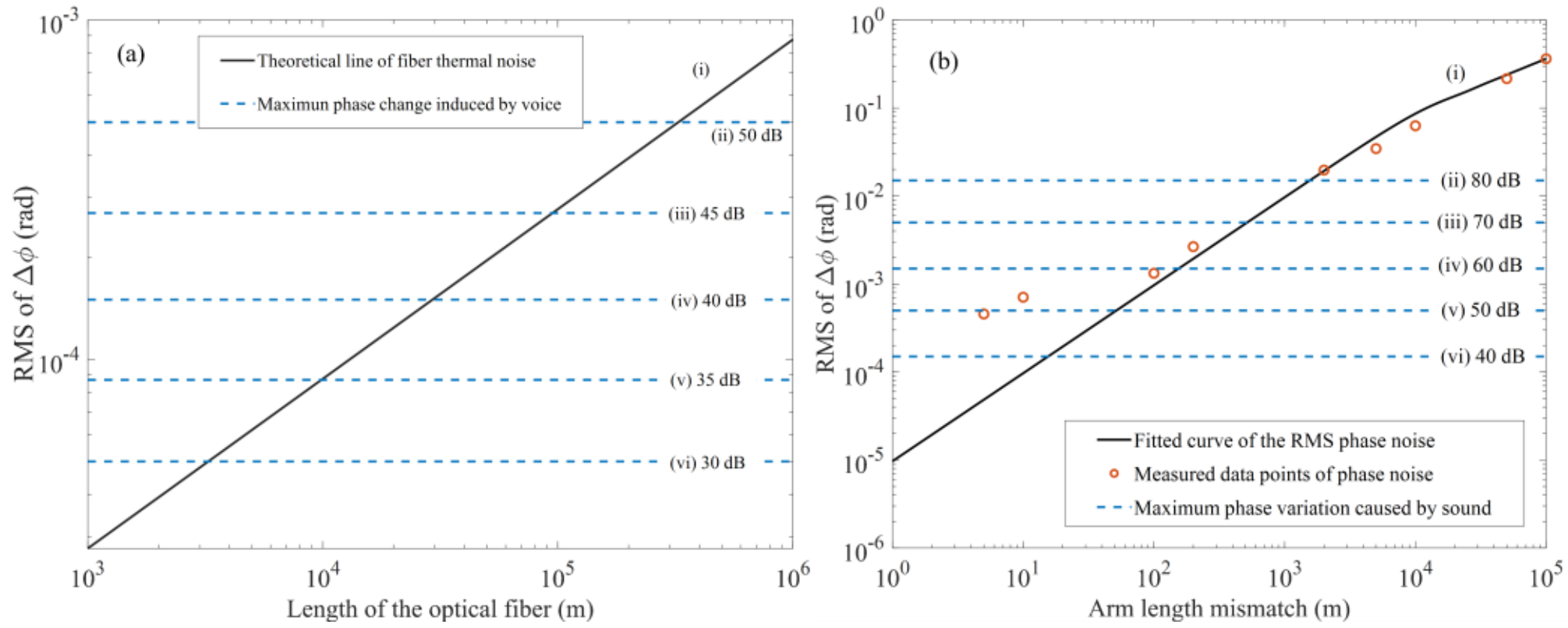

Fig. 2. Phase noise in the eavesdropping system. (a) Line i: theoretical fiber thermal noise at different fiber lengths. Line ii ~ vi: maximum phase fluctuations in the sensing fiber caused by 50, 45, 40, 35, and 30 dB sound. (b) Relationship between RMS phase fluctuations and length mismatches of interferometer arms. Line i: fitted curve of phase fluctuations. Line ii ~ vi: phase variations in the sensing fiber caused by 80, 70, 60, 50, and 40 dB sound. Red circle: measured data points of phase noise.

The relationship between the RMS of thermal noise and the fiber length is shown in Fig. 2(a). The phase change caused by sound pressure on the sensing fiber is presented as well, which is measured and estimated from the results of the experiment. The level of thermal noise is proportional to the fiber length, which determines the detecting limit of the system. It can be seen from Fig. 2(a) that to detect 30 dB sound, the length of the detecting arm is limited to around 3 kilometers. Considering that human voice is normally much louder than 30 dB, the limitation here is quite loose, which means eavesdropping will not be seriously disturbed by fiber thermal noise.

The laser frequency fluctuates with time because of spontaneous emission [23, 24]. Therefore, the phase difference between $\varphi_{Laser}(t)$ and $\varphi_{Laser}(t-\tau_0)$ varies with time and induces noise in the gathered signal. The RMS of laser source-caused phase variation in the gathered signal can be expressed as:

$$\sqrt{\left\langle \Delta\phi_{\tau_0}^2(t) \right\rangle} = \left( \left\langle \left[ \varphi_{Laser}(t) - \varphi_{Laser}(t-\tau_0) \right]^2 \right\rangle \right)^{1/2}, \tag{7}$$

The PSD of this term can be derived from the frequency fluctuation characteristics of the laser source.

The PSD of the frequency fluctuations of the laser source can be written as [25]:

$$S_{\dot{\phi}}(f) = S_0 + k/f, \tag{8}$$

where $S_0$ and $k$ are constants related to the linewidth of the laser source. The phase difference from time $t-\tau_0$ to $t$ can be expressed as the integral of the frequency:

$$\Delta\phi_{\tau_0}(t) = \int_{t-\tau_0}^{t} \dot{\phi}(s)\,ds. \tag{9}$$

Therefore, the PSD of laser source-caused phase variation in the gathered signal can be written as [26]:

$$S_{\Delta\phi_{\tau_0}}(f) = \frac{\sin^2(\pi f \tau_0)}{(\pi f)^2}\left( S_0 + \frac{k}{f} \right). \tag{10}$$

It is clear from equation (10) that laser source-caused phase variation is determined by two major factors. One is the linewidth of the laser source, which means choosing a narrow linewidth laser source will reduce the background noise. The other is the time delay between two arms of the interferometer. Normally, the mismatching length between two arms is no more than several kilometers, with a time delay of no more than 10 μs. The product of $\tau_0$ and $f$ (less than 10 kHz) is much less than one. Equation (10) can be simplified as:

$$S_{\Delta\phi_{\tau_0}}(f) = \tau_0^2\left( S_0 + \frac{k}{f} \right). \tag{11}$$

The PSD of laser source-caused phase noise is proportional to the square of $\tau_0$. Therefore, making the interferometer a balanced one could effectively reduce the background noise, and improve the SNR of the signal.

The RMS of laser source-caused phase variation can be expressed as:

$$\sqrt{\left\langle \Delta\phi_{\tau_0}^2(t) \right\rangle} = \left( \int_{f_L}^{f_H} S_{\Delta\phi_{\tau_0}}(f)\,df \right)^{1/2} \tag{12}$$

where $f_H$ and $f_L$ are consistent with previous interpretation. The relationship between RMS phase variation and the arm length mismatch of the interferometer is shown in Fig. 2(b). The value of $S_0$ and $k$ is fitted by the acquired data. The fitted curve is basically consistent with the actual measured points, while two points measured at 5 meters and 10 meters have higher phase noise levels than the fitted line. It is mainly caused by environmental noise in the lab and the circuit noise from DAQ. When the arm mismatch increases, PSD of laser source-caused phase variation raises quadratically and laser source-caused phase noise becomes the most important noise immediately. Therefore, it is necessary to balance the arm length of the interferometer while deploying the eavesdropping system.

## 4. Results

In the experiment, we measured the length of the two arms by OTDR and balanced the interferometer. The arm length mismatch is less than 100 meters. Theoretically, the noise of the system should be relatively small at this time, and the detecting limit is around 60 dB according to Fig. 2(b).

During our experiment, the whole system is influenced by the environmental noise, which is mainly introduced by the air conditioner and fans of the devices in the lab. The noise is around 55 dB and is concentrated in the low-frequency range (below 500 Hz). We have taken some measures to reduce its effect. However, it still introduces high phase noise at low frequencies. Therefore, the heterodyne signal is high-pass filtered with a 500 Hz cut-off frequency to remove these impacts. Although part of the sound signal has been removed together with the noises in this process, the remaining part can still perform well in the waveform. Actually, in real eavesdropping scenes, this environmental noise will be lower than the lab test and won't influence the eavesdropping too much. The reason is that fibers connecting the eavesdropper and the indoor segment will be relatively shorter, protected by hard layer, and placed in the fiber pipes. That's the reason why we don't take it into account in theoretical analysis. The experimental results are shown in Fig. 3.

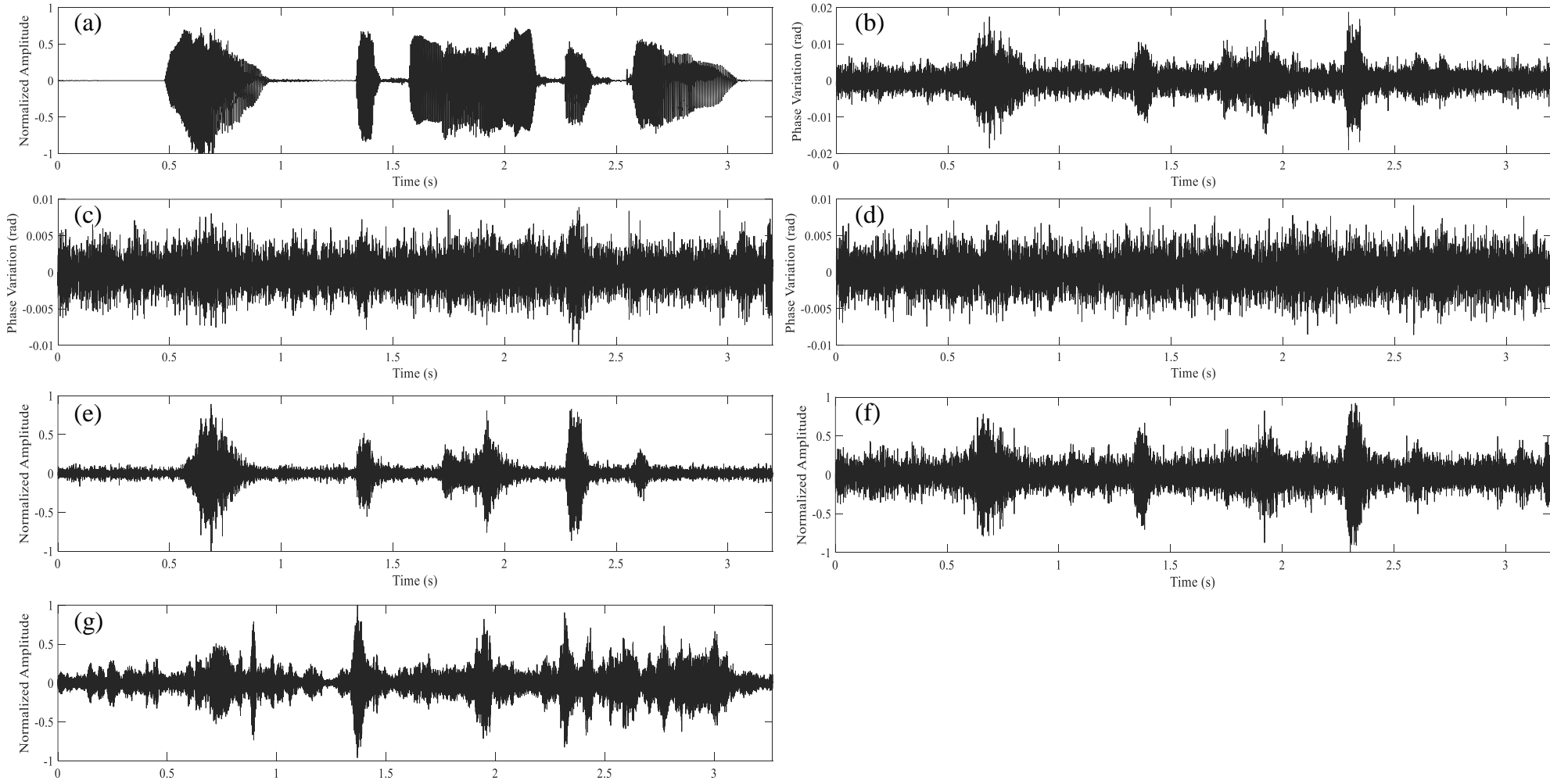

Fig. 3. Experimental results in different conditions. (a) The waveform of source audio (see also Visualization 1). (b) Phase variation of the heterodyne signal when applying 70~80dB sound. (c) Phase variation of the heterodyne signal when applying 60~70dB sound. (d) Phase variation of the heterodyne signal when applying 50~60dB sound. (e) Audio signal after noise reduction and further filter from (b) (see also Visualization 2). (f) Audio signal after noise reduction and further filter from (c) (see also Visualization 3). (g) Audio signal after noise reduction and further filter from (d) (see also Visualization 4).

Fig. 3(a) shows the waveform of the audio source, the content of which is "it's nine-fifteen". The audio file is available in Visualization 1. Fig. 3(b)~(d) shows the phase variation of heterodyne signal collected when the sound acting on the sensing fiber is 70~80 dB, 60~70 dB and 50~60 dB, respectively. The waveform of the 70~80 dB one can be seen clearly, while the 50~60 dB one is completely submerged in the background noise. To show that we have still detected the signal at 50~60 dB, speech enhancement methods are applied to remove part of the background noise. Considering that the noise is a stationary random signal, the spectrum of noise is statistically constant between frames. Therefore, it is possible to estimate the power spectrum of noise in silent frames and subtracted its spectrum from frames containing signals. This spectrum subtract method [27, 28] is simple but effective.

Fig. 3(e)~(g) shows the enhancement result of Fig. 3(b)~ (d). In Fig. 3(e) and Fig. 3(f), several obvious audio waveforms are retained. The audio file of Fig.3(e) and Fig. 3(f) is available in Visualization 2 and Visualization 3, which can still be heard clearly. In Fig. 3(g), the several main audio waveforms in Fig. 3(a) can also be seen after eliminating partial noise and the waveform is visible, and the audio file is available in Visualization 4. Although the recognizability of voice information has been affected, advanced speech enhancement methods such as OMLSA [29] and recognition methods based on deep neural networks [30], may be used to recover voice information from the audio file. However, this is beyond the scope of this paper.

## 5. Discussion

The eavesdropping method shown in this paper requires complicated equipment and strict conditions. However, the secret stealing behavior is always done regardless of cost. Therefore, for some occasions with strict confidentiality requirements, leakage through communication optical fiber should be prevented. Ref. [18] invents a device for detecting whether it is being eavesdropped in the optical fiber network. In this section, several ways to prevent eavesdropping are given, some of which are low-cost and easy to implement.

The phase change in a section of optical fiber caused by external pressure can be described as [31, 32]:

$$\frac{\Delta\varphi}{\varphi} = \varepsilon_z - \frac{n^2}{2}\left(\left(P_{11} + P_{12}\right)\varepsilon_r + P_{12}\varepsilon_z\right), \tag{13}$$

where $\varepsilon_z$ is the axial strain and $\varepsilon_r$ is the radical strain of the fiber core, which is determined by the variation of pressure and the structure of the optical fiber. $P_{11}$ and $P_{12}$ are the Pockels coefficient of the core. $\varphi$ is the phase change of the light wave through this section of fiber. The method of solving $\varepsilon_z$ and $\varepsilon_r$ can be found in [33, 34]. It can be seen from equation (12) that the phase change is a combined effect of both the change of optical path length induced by fiber length and refraction index induced by strain. These two factors have opposite effects on the change of phase. Therefore, it is possible to design pressure-insensitive fiber by choosing appropriate coat materials and thickness.

It can also be seen from equation (12) that under the same conditions, the phase change caused by pressure variation is proportional to the optical path length. Therefore, decreasing the indoor fiber length is a simple method to reduce the risk of being eavesdropped on. What's more, previous studies have shown that the phase change caused by the same pressure is inversely proportional to the bulk modulus of the optical fiber cables. Using high bulk modulus materials, such as metal and glass, as the coat of fiber could effectively increase the bulk modulus of the whole fiber and reduce the phase change introduced by sound.

Changing the fiber adapter connecting to the modem from a flat PC end face to an angled APC end face can reduce the echo, which is also a practical method to avoid the risk of being eavesdropped. However, in current digital communication systems which do not care about reflected light, PC adaptors are still the most popular on account of their low cost.

Figure. 4 shows the experimental results of reducing fiber acoustic sensitivity. Figure. 4(a) shows the spectrum of the audio source and Fig. 4(b) shows the spectrum of sound signal gathered at 60~70 dB with a 3-meter sensing fiber. In Fig. 4(c), the length of the sensing fiber is changed from 3-meter to 1-meter. In Fig. 4(d), the sensing fiber is tied with a 1 mm diameter steel wire to improve its bulk modulus. It is obvious that the voice signal in Fig. 4(b) vanishes in Fig. 4(c) and Fig. 4(d), which shows the effectiveness of these measures. The comprehensive use of these methods can simply increase the cost of eavesdropping, so as to effectively avoid its risk.

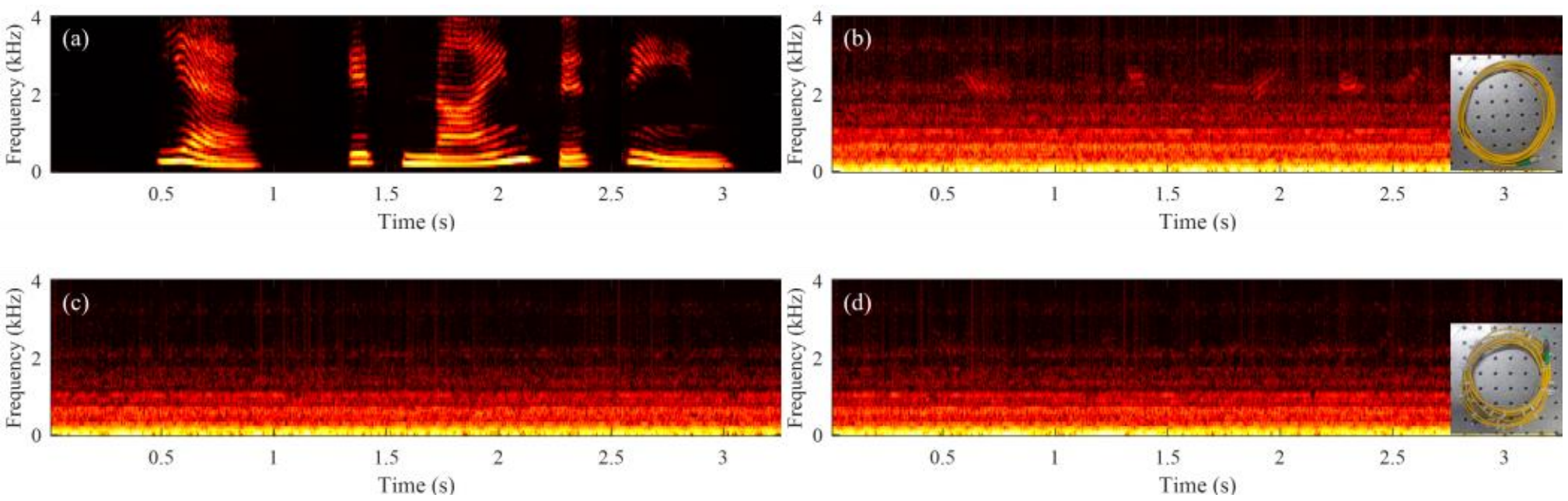

Fig. 4. The spectrum of audio signals. (a) Audio source. (b) Audio signal gathered through 3m fiber with 70dB sound. (c) Audio signal gathered through 1m fiber with 70dB sound. (d) Audio signal gathered through 3m fiber bonded to a 1 mm diameter steel wire with 70dB sound.

## 6. Conclusion

In this paper, we present an eavesdropping method using indoor optical fiber. We show that the background noise level can be reduced by balancing the interferometer arms and give the detection limit of the system. Experimental results show that it is possible to detect and recover sound signals through communication fiber networks secretly. Several measures to prevent eavesdropping are also discussed.

**Funding.** National Natural Science Foundation of China (62171249, 61971259), and Tsinghua University Initiative Scientific Research Program.

**Disclosures.** The authors declare no conflicts of interest.

**Data availability.** Data underlying the results presented in this paper are not publicly available at this time but may be obtained from the authors upon reasonable request.